\documentclass[journal,article,accept,pdftex,moreauthors]{Definitions/mdpi}
\firstpage{1}
\pubvolume{1}
\issuenum{1}
\articlenumber{0}
\pubyear{2026}
\copyrightyear{2026}
\datereceived{ }
\daterevised{ } 
\dateaccepted{ }
\datepublished{ }

\usepackage{graphicx}%
\usepackage{multirow}%
\usepackage{amsmath}%
\usepackage{amsthm}%
\usepackage{mathrsfs}%
\usepackage[title]{appendix}%
\usepackage{xcolor}%
\usepackage{textcomp}%
\usepackage{manyfoot}%
\usepackage{booktabs}%
\usepackage{algorithm}%
\usepackage{algorithmicx}%
\usepackage{algpseudocode}%
\usepackage{rotating}%
\usepackage{adjustbox}%

\usepackage{listings}%
\lstdefinelanguage{Tcl}{
  morekeywords={proc, puts, return, set, if, else, elseif, error, eq, ne, namespace, eval, exec},
  sensitive=true,
  morecomment=[l]{//},
  morecomment=[s]{/*}{*/},
  morestring=[b]",
}

\Title{HBS - Hardware Build System: Characterizing and comparing direct-Tcl and indirect-abstract approaches for hardware build systems}

\Author{Michał Kruszewski $^{1}$\orcidA{}}

\AuthorNames{Michał Kruszewski}

\address{%
$^{1}$ \quad Warsaw University of Technology, Faculty of Electronics and Information Technology, Institute of Electronic Systems, Nowowiejska 15/19, 00-650 Warsaw, Poland}

\corres{Correspondence: michal.kruszewski@pw.edu.pl}

\abstract{
Build systems become an indispensable part of the software implementation and deployment process.
New programming languages are released with the build system integrated into the language tools, for example, Go, Rust, or Zig.
However, in the hardware description domain, no official build systems have been released with the predominant Hardware Description Languages (HDL) such as VHDL or SystemVerilog.
Moreover, hardware design projects are often multilingual.
The paper characterizes and compares two common approaches for hardware build system implementations.
The first one, the direct-Tcl approach, in which the build system code is executed directly by the EDA tool during the design build flow.
The second one, the indirect-abstract approach, in which the build system produces a Tcl script, which is later run by a proper EDA tool.
As none of the existing direct-Tcl build systems was close to the indirect-abstract build systems in terms of supported functionalities, the paper also presents a new direct-Tcl hardware build system called HBS.
The implemented build system was used as a representative of direct-Tcl build systems in comparison with indirect-abstract build systems.
}


\begin{document}

\section{Introduction}
\label{sec:approaches}

Implementation of any complex software or hardware description design involves not only code writing.
Equally important is defining metadata used for configuration, dependency management, code generation, testing, and deployment purposes.
The metadata is read by software tools, which, in turn, generate final binary or bitstream files or perform additional tasks, such as unit tests.
Such programs are often referred to as build systems.
Some build systems are split into multiple smaller programs, and some have a monolithic structure.
An introduction to the build system concept and principles can be found in \cite{software_build_systems_principles_and_experience, build_systems_a_la_carte_theory_and_practice}.

In the software domain, there are multiple well-established build systems, such as GNU Make \cite{gnu_make}, CMake \cite{cmake}, Ninja \cite{ninja}, or Bazel \cite{woset2021_bazel, bazel}.
Some of them were implemented in the early days of programming.
For example, GNU Make was first released in 1988 as part of the GNU Project.
It was implemented to provide a free software alternative to Unix Make, which was first implemented in 1976 by Stuart Feldman at Bell Labs \cite{Feldman:1979:MPM}.
Modern programming languages, such as Go, Rust, or Zig, include build systems in their toolchains.
The Zig build system can even be used as a build system for C/C++ projects.

In the hardware domain, the situation looks different.
The earliest Electronic Design Automation (EDA) tools were implemented as proprietary programs.
EDA vendors did not (and still do not) cooperate to create a standard (or at least similar) user interface for tool interaction.
The main common point among various EDA tools is the use of the Tcl language as the interpreter for interacting with the tools.
However, the tools provide custom commands and parameters even for primary, common tasks, such as adding hardware description files to a project design.
For example, to add a VHDL file in AMD Vivado, one can use specific \texttt{read\_vhdl <path\_to\_vhdl\_file>} or generic \texttt{add\_files <path\_to\_vhdl\_file>} commands.
To add a VHDL file in Intel/Altera Quartus, one has to use \texttt{set\_global\_assignment -name VHDL\_FILE <path\_to\_vhdl\_file>} command.
EDA tools have abundant differences, making implementing a build system for hardware projects harder than for software.
Despite this, there have been (and still are) attempts to implement build systems better suited to hardware projects.

The main goal of the paper is to characterize and compare two existing approaches for hardware build system implementation, the direct-Tcl and indirect-abstract.
The paper also partially describes the new hardware build system (HBS), which follows the direct-Tcl approach.
The system was designed and implemented to serve as a comparison representative.

The article's structure is as follows.
Section \ref{sec:approaches} describes two commonly used approaches for implementing hardware build systems.
Section \ref{sec:why} explains why HBS was implemented.
Section \ref{sec:requirements} describes requirements imposed on the HBS implementation and user interface.
Section \ref{sec:architecture} briefly introduces the internal architecture of HBS.
Section \ref{sec:comparison} compares various aspects of the direct-Tcl and indirect-abstract approaches.
Section \ref{sec:summary} summarizes the work.

\section{Hardware build system approaches}
\label{sec:approaches}

There exist at least more than ten build systems targeting the hardware domain.
Some of them are vendor-tool-specific, while others aim to be generic, entirely abstracting away the tool used for synthesis or implementation.
By looking at the architecture (analyzing how a tool internally works), existing hardware build systems can be classified into two approaches:
\begin{enumerate}
  \item
    Direct-Tcl approach - the build system code is executed directly by the EDA tool during the design build flow.
    The build system code and the build code reside in the same domain.
    The build system code has access to the build code and vice versa.
  \item
    Indirect abstract approach - the build system code produces a Tcl script, which is only later run by a proper EDA tool.
    The build system code and the build code reside in different domains.
    The build system code generates the build code, but it does not have access to it during the design build flow.
    The build code is not aware of the build system code and has no access to it.
\end{enumerate}

\subsection{Direct-Tcl approach}

In the direct-Tcl approach, the user directly writes Tcl code and uses the EDA tools' Tcl interface.
The idea seems obvious.
Since EDA tools already utilize Tcl as a user interface, a build system should also provide users with a Tcl interface.

Usually, the direct-Tcl approach build systems implement some wrapper commands that the user should use for primary tasks, such as, file addition, library setting, or generic/parameter setting.
However, in the case of more specific tasks, the user is supposed to call the EDA tool's specific commands explicitly.
It is also up to the user to maintain build compatibility across various EDA tools when build requirements are more complex, for example, when constraints depend on synthesis results.
It is often achieved with tool guard clauses.
For example, an if statement block calling a specific set of commands only if the currently executed EDA tool is specific.
Representatives of the direct-Tcl approach are, for example, fpga-vbs \cite{fpga_vbs_git}, vextproj \cite{vextproj_git, vextproj_spie}, and OSVVM-Scripts \cite{osvvm_scripts_git}.

Easily executing arbitrary, custom EDA tool commands is the most significant advantage of the direct-Tcl approach.
As a Tcl interpreter embedded in an EDA tool executes the build system code, a user can call custom commands almost anywhere.

The direct-Tcl approach is not free of drawbacks.
Some of them are inherent, and others result from the fact that direct-Tcl build systems development was stopped at some stage by their maintainers.

The first inherent drawback is that Tcl is considered an unfriendly language compared to, for example, Python.
The argument is often raised by junior hardware engineers and has a very subjective nature.
However, Tcl was chosen as the EDA scripting language because it was very popular when EDA tools were developed \cite{scripting_for_eda}.
The second inherent drawback is that Tcl is not a popular language nowadays.
Although there are no physical obstacles to change this trend, it is doubtful that suddenly more developers will start using Tcl instead of, for example, Python as an interpreted language.
This is why this drawback is also treated as inherent.

The second group of direct-Tcl build systems drawbacks results from stopping development at a certain point.
There is no direct-Tcl build system that can be used to examine the limits and bottlenecks of this idea.
For example, most of the direct-Tcl build systems support only one EDA tool.
OSVVM-Scripts is an exception.
However, it can be used only for simulation.
It does not support any synthesis or implementation tools.
More example drawbacks common to the direct-Tcl build systems include:
\begin{enumerate}
  \item No dependency graph generation capability.
  \item Lack or limited capability of automatic cores, targets, and testbenches detection.
  \item Lack of automatic parallel testbench running.
\end{enumerate}

The HBS was designed and implemented to examine the limits and bottlenecks of the direct-Tcl approach.
To the author's knowledge, there is no direct-Tcl approach hardware build system, supporting multiple EDA tools, with features and capabilities similar to those provided by the HBS.

\subsection{Indirect-abstract approach}

The indirect-abstract approach is much more popular than the direct-Tcl approach.
In this approach, the user declares various properties and configurations rather than writing Tcl code or calling EDA tools’ Tcl commands directly.
The main characteristic of the indirect-abstract approach is the execution of the build system code before the actual EDA tool is run.
The result of the build system execution is a Tcl, Makefile, or a shell script, which is only later run by the specified EDA tool.

The most popular hardware build system utilizing the indirect-abstract approach is probably FuseSoc \cite{fusesoc, fusesoc_git}.
Other popular build systems include
SiliconCompiler \cite{siliconcompiler, siliconcompiler_git},
HAMMER \cite{hammer, hammer_git},
Hog \cite{hog, hog_git},
hdlmake \cite{hdlmake_web},
orbit \cite{orbit_git},
bender \cite{bender_git},
bazel\_rules\_hdl \cite{bazel_rules_hdl_git},
flgen \cite{flgen_git},
and Blockwork \cite{blockwork_git}.

The indirect-abstract hardware build systems can be subdivided into the following three subcategories based on the language and syntax they utilize:
\begin{enumerate}
  \item
    Declarative - the design build flow is described in a fully declarative way utilizing markup or data serialization languages, for example, YAML.
    An example of the declarative indirect-abstract build system is FuseSoc.
  \item
    Programmable - the design build flow is described using a fully functional programming language, for example, Python.
    An example of the programmable indirect-abstract build system is SiliconCompiler.
  \item
    Mixed - the design build flow is partially described using the declarative way and partially programmable.
    An example of the mixed indirect-abstract build system is Hog.
\end{enumerate}

The language and syntax used for describing build flow vary between different build systems.
Some use simple data serialization formats such as YAML (for example, FuseSoc or bender) or TOML (for example, orbit).
Some come with a domain-specific language (for example, flgen), or use Python (for example, SiliconCompiler, HAMMER) and Makefiles (for example, Hdlmake).
Finally, some use a mix of more than one (for example, Hog).
Especially using Python as a programming language is powerful, as users can use all functionalities that come with the language.

The most significant advantage of the indirect-abstract approach is the low entry threshold.
The syntax of languages such as YAML, TOML, or Python is simple, which results in high legibility.
Even junior engineers without extra training can understand it.
Listing \ref{lst:fusesoc_edge_detector} presents a very basic core definition in the FuseSoc build system.
The first line specifies the version of the description format.
The ``name'' key specifies the vendor, library, name, and version (VLNV) identifier specified in the IP-XACT User Guide \cite{ip_xact_user_guide}.
The remaining part is relatively self-explanatory.

\begin{lstlisting}[
label=lst:fusesoc_edge_detector,
caption=Example edge detector core definition in FuseSoc build system.
]
CAPI=2:

name: vendor:library:edge_detector:1.0

filesets:
  src:
    files:
      - src/edge_detector.vhd
    file_type: vhdlSource-2008
    logical_name: hdl_library_name

targets:
  default:
    filesets:
      - src
\end{lstlisting}

The indirect-abstract approach is based on the concept that the interface to EDA tools is hidden under some common abstraction layer.
Whether this is an advantage or a disadvantage is debatable.
Providing a common abstract layer means that users do not have to learn the interface of multiple EDA tools.
Moreover, in theory, changing the EDA tool for the compilation process should be limited to changing the EDA tool declaration solely.
However, this is somewhat of a utopian view.
As the practice shows, the common abstraction layer leads to the following common difficulties:
\begin{enumerate}
  \item
    Missing feature implementation for some EDA tools.
    Let us suppose the abstraction layer declares feature \emph{F}, and the build system claims support for tools \emph{T1} and \emph{T2}.
    Although \emph{T1} might provide the implementation for \emph{F}, \emph{T2} might be missing the implementation for this particular feature.
    In such a case, the user must add support for this feature to the build system.
    This task might not be trivial.
    The user must understand the internals of the build system, prepare a pull request, wait for a review, and potentially reiterate the implementation.
    Yet another approach is to modify automatically generated build scripts before a compilation.
    However, this is against the whole idea of build systems.
    A direct-Tcl approach is free of these difficulties because the user can directly insert custom Tcl commands.
    There are no intermediate abstraction layers.
 \item
    Simple tasks, such as scoping constraints to a particular module, sometimes require verbose solutions.
    The solution is not only more verbose but also more complex to understand.
    What can be solved in one or two lines of Tcl code requires more than one file.
    An example is discussed in subsection \ref{subsec:constr-handling}.
\end{enumerate}

All indirect-abstract hardware build systems allow ``injecting'' arbitrary custom Tcl commands into the automatically generated scripts.
However, these commands can not be inserted into arbitrary places.
Moreover, inserting a single-line Tcl command often requires more than one line of code or even multiple files.
This can be treated as proof that the indirect-abstract approach is not a golden bullet.
The idea of a common abstraction layer has significant drawbacks that no one has yet found a way to fix.

Making a truly abstract hardware build system is highly unlikely in the near future.
The main problem with the common abstraction layer is the fact that EDA tools were implemented before such a layer was proposed.
Declaring a common interface for existing tools is a much harder task than implementing tools that conform to a formal interface.

\section{Why Hardware Build System was implemented}
\label{sec:why}

Characterizing and comparing direct-Tcl and indirect-abstract approaches for hardware build systems theoretically did not require HBS implementation.
However, none of the existing direct-Tcl build systems was functionally close to the existing indirect-abstract build systems.
In particular, none of the existing direct-Tcl build systems offered the following functionalities:
\begin{enumerate}
  \item support for multiple synthesis EDA tools,
  \item listing automatically detected cores and targets,
  \item automatic testbench targets detection,
  \item parallel testbench running,
  \item dependency graph generation.
\end{enumerate}

The author wanted to implement a direct-Tcl build system that would functionally adhere to the existing indirect-abstract build systems.
It had to be empirically verified if it is possible to have features of indirect-abstract build systems in a direct-Tcl build system.

The implemented build system (HBS) serves as a representative of the direct-Tcl approach for the comparison with indirect-abstract build systems.

\section{Hardware Build System requirements}
\label{sec:requirements}

The main goal of the HBS implementation was to push the direct-Tcl approach to its limits, examine its bottlenecks, and evaluate the concept of direct-Tcl build system by comparing it with indirect-abstract approaches.
However, it does not mean that HBS is solely a prototype and was designed without thoughtful requirements.
This section provides a summary of HBS major requirements with extra explanation.

\subsection{Simplicity and ease of use}
\label{lbl:req_simplicity}

The overhead of build system implementation and maintenance is often underrated \cite{changes_to_build_systems_2024, coevo_source_code_and_build_system_2009}.
McIntosh in \cite{build_system_maintenance_2011} reports that the build system can account for up to 31\% of the code files in a project.
Moreover, up to 27\% of development tasks that change the source code also require build system maintenance.
The study was carried out on software projects, not on hardware designs.
However, due to the complexity of hardware design, it seems reasonable to assume comparable shares.

Nejati et al. in \cite{code_review_of_build_system_specifications_2023} report that the complexity of build system specifications is one of the top three factors complicating the review of build speciﬁcations.

The primary goal of the hardware build system proposed in the paper is simplicity.
Contemporary hardware build systems utilizing the indirect-abstract approach often offer simplicity only at the initial stage.
However, as the project's complexity grows, so does the complexity of using the build system.

\subsection{Minimal user interface}

Xia et al. \cite{empirical_study_of_bugs_2013} report that even up to 21.35\% of the bugs in software build systems belong to the external interface category.
This means that one-fifth of build system bugs result from incorrect usage of the user interface or different behavior of the build system with different tools or platforms.

Fixing build system external interface bugs requires applying patches in the build system itself.
Such a task is relatively time-consuming and might not be simple to implement.
The build system user must report the problem to the build system developers or fix the bug themselves.
However, this implies that the build system user is distracted from their own work.
Moreover, the fix is not applied immediately.
This means that the user must also find a temporary workaround.
This, in turn, leads to a technical debt \cite{productivity_loss_due_to_technical_debt_2019, technical_debt_cripples_2018}.

A minimal user interface has one more advantage.
It allows developers to customize build systems to the project's needs.
Each project is in some way unique and might require a particular approach to solve problems related to the design build.
Moreover, projects usually evolve, for example, due to customer demand.
Dynamic tailoring to the project requirements is a desired feature of modern build systems \cite{future_of_build_systems_2018}.
Similar observations were made by Neitsh et al. in \cite{build_system_multilanguage_2012}.
Authors report that concerns from the application and implementation domains may ``leak'' into the build model.
The build system should be adaptable enough to provide ways to respond to these concerns.
Too rigid user interfaces limit such capabilities.

\subsection{Low number of dependencies}
\label{lbl:req_low_deps}

Xiao et al. \cite{technical_debt_in_build_systems_2022} report  that dependency management is one of the top three factors contributing to the introduction of technical debt in build systems.
Such a finding is not surprising.
Dependencies often break in unexpected moments.
In such a case, long-term quality is sacrificed for the short-term goals.
Getting a project to compile successfully is more important than fixing broken build system dependencies.

One of the goals of the HBS is to minimize the probability of user distraction caused by broken dependencies.
This implies that the number of dependencies should be kept as low as possible.

Moreover, a low number of dependencies potentially makes the build system more appealing for use in military designs.
This is because any external dependency must be verified against potential vulnerabilities.

\subsection{Ease of calling arbitrary programs in arbitrary places}
\label{lbl:req_ease_of_calling}

The ease of calling arbitrary programs in arbitrary places results from two use cases.

The first use case originates from the complexity of designing hardware systems.
To reduce manual work, engineers often automatically generate parts of the design.
For example, register files are automatically generated from SystemRDL \cite{systemrdl}.
In such a case, there must be a way to call an external program that generates hardware description of the register components.

Existing indirect-abstract hardware build systems allow calling external programs.
However, they often require users to define a transitional so-called generator, which increases complexity and reduces legibility.
Moreover, they usually do not allow calling external programs in arbitrary places.
This is further discussed in subsection \ref{subsec:external-program-calling}.

The second use case includes support for EDA tools without the Tcl interface.
Such tools are often open source simulators, for example, nvc \cite{nvc_git}, or verilator \cite{verilator_git}.
These tools have interfaces similar to the standard software compiler.
The user simply calls programs with proper arguments and proper order in a shell.
Being able to call arbitrary programs in arbitrary places is required to support these kinds of tools without an extra script or Makefile generation.

\subsection{Independence from version control system}

Some hardware build systems are tightly coupled with a particular version control system.
For example, Hog is intended to be used only with git.
However, build systems and version control systems are conceptually orthogonal entities.
Imposing a particular version control system limits user choice.
Moreover, it potentially makes incorporating a new build system for some existing projects impossible, as they might, for example, utilize Subversion instead of git.
\section{Hardware Build System architecture}
\label{sec:architecture}

The HBS official user manual is available in its repository \cite{hbs_git}.
The purpose of this section is to briefly describe the architecture of HBS so that the reader does not have to study the user manual.

The HBS was implemented to evaluate the limitations of the direct-Tcl approach.
Nevertheless, implementing a hardware build system using this approach may be a viable strategy for the following reasons:
\begin{enumerate}
  \item
    Implementing a hardware build system in Tcl allows the execution of build system code during the EDA tool flow.
    This, in turn, gives direct access to all EDA tool custom commands.
    Moreover, these custom commands can be evaluated in arbitrary places.
    This aspect satisfies requirements \ref{lbl:req_simplicity} and \ref{lbl:req_ease_of_calling}.
    The complexity of existing indirect-abstract build systems for less common tasks results from executing build system code before the EDA tool flow.
    The build system only prepares scripts that are later run by EDA tools.
  \item
    If the EDA tool provides the Tcl interface, then the Tcl shell is provided by the EDA tool vendor.
    The shell is installed during the installation of vendor tools.
    This implies that, in some cases, the build system user does not even have to install additional programs.
    This aspect satisfies requirement \ref{lbl:req_low_deps}.
  \item
    Executing arbitrary programs in arbitrary places in Tcl is very straightforward.
    There is a dedicated \texttt{exec} command for invoking subprocesses.
    If executing a subprocess requires prior dynamic arguments evaluation, the \texttt{exec} command call must be prepended with the \texttt{eval} command.
    Even in Python, invoking a subprocess is not so simple.
    This Tcl feature satisfies requirement \ref{lbl:req_ease_of_calling}.
\end{enumerate}

\subsection{Cores and cores detection}

When the user executes \texttt{hbs}, all directories under the working directory are recursively scanned to discover all \texttt{.hbs} files (including symbolic links).
Files with the \texttt{.hbs} extension are regular Tcl files that are sourced by the \texttt{hbs} script.
However, before sourcing \texttt{.hbs} files, the file list is sorted so that scripts with shorter path depth are sourced first.
For example, let us assume the following three \texttt{.hbs} files were found:
\begin{itemize}
  \item \texttt{a/b/c/foo.hbs},
  \item \texttt{d/bar.hbs},
  \item \texttt{e/f/zaz.hbs}.
\end{itemize}
Then, they would be sourced in the following order:
\begin{enumerate}
  \item \texttt{d/bar.hbs},
  \item \texttt{e/f/zaz.hbs},
  \item \texttt{a/b/c/foo.hbs}.
\end{enumerate}
Such an approach allows controlling when custom symbols (Tcl variables and procedures) are ready to use.
For example, if the user has a custom procedure used in multiple \texttt{.hbs} files, then the user can create separate \texttt{utils.hbs} file containing utility procedures, and place it in the project root directory.
Within \texttt{.hbs} files, the user usually defines cores and targets, though they are free to include any valid Tcl code.

Listing \ref{lst:flip-flop-hbs} presents a very basic flip-flop core definition.
The \texttt{flip-flop} core has a single target named \texttt{src}.
The core consists of a single VHDL file.
\begin{lstlisting}[
  language=Tcl,
  label=lst:flip-flop-hbs,
  caption=Example flip-flop core definition in HBS.
]
namespace eval flip-flop {
  proc src {} {
    hbs::AddFile flip-flop.vhd
  }
  hbs::Register
}
\end{lstlisting}

To register a core, the user must explicitly call the \texttt{hbs::Register} procedure at the end of the core namespace.
Such a mechanism helps distinguish regular Tcl namespaces from those representing core definitions.

Each core is identified by its unique path.
The core path is equivalent to the namespace path in which \texttt{hbs::Register} is called.
Using the namespace path as the core path gives the following possibilities:
\begin{enumerate}
  \item
    The user can easily stick to the VLNV identifiers if required.
    This is presented in Listing \ref{lst:flip-flop-hbs-vlnv}.
    In this case, the flip-flop core path is \texttt{vendor::library::flip-flop::1.0}.
  \item
    The user can define arbitrary deep core paths (limited by the Tcl shell).
    This is presented in Listing \ref{lst:flip-flop-hbs-deep-path}.
    In this case, the core path consists of 7 parts.
  \item
    The user can nest namespaces to imitate the structure of libraries and packages.
    This is presented in Listing \ref{lst:flip-flops-hbs}.
    Three flip-flop cores are defined in the snippet.
    Listing \ref{lst:list-cores-flip-flops} presents output for listing flip-flop cores.
\end{enumerate}

\begin{lstlisting}[
  language=Tcl,
  label=lst:flip-flop-hbs-vlnv,
  caption=Example flip-flop core definition with VLNV core path.
]
namespace eval vendor::library::flip-flop::1.0 {
  proc src {} {
    hbs::AddFile flip-flop.vhd
  }
  hbs::Register
}
\end{lstlisting}

\begin{lstlisting}[
  language=Tcl,
  label=lst:flip-flop-hbs-deep-path,
  caption=Example flip-flop core definition with 7 core path parts.
]
namespace eval a::b::c::d::e::f::flip-flop {
  proc src {} {
    hbs::AddFile flip-flop.vhd
  }
  hbs::Register
}
\end{lstlisting}

\begin{lstlisting}[
  language=Tcl,
  label=lst:flip-flops-hbs,
  caption=An example \texttt{.hbs} file with three flip-flops core definitions
]
namespace eval lib {
  namespace eval pkg1 {
    namespace eval d-flip-flop {
      proc src {} {
        hbs::AddFile d-flip-flop.vhd
      }
      hbs::Register
    }
    namespace eval t-flip-flop {
      proc src {} {
        hbs::AddFile t-flip-flop.vhd
      }
      hbs::Register
    }
  }
  namespace eval pkg2 {
    namespace eval jk-flip-flop {
      proc src {} {
        hbs::AddFile jk-flip-flop.vhd
      }
      hbs::Register
    }
  }
}
\end{lstlisting}

\begin{lstlisting}[
  label=lst:list-cores-flip-flops,
  caption=Listing flip-flop cores output.
]
[user@host tmp]$ hbs ls-cores
lib::pkg1::d-flip-flop
lib::pkg1::t-flip-flop
lib::pkg2::jk-flip-flop
\end{lstlisting}

\subsection{Targets and targets detection}

The HBS automatically detects targets.
Targets are all Tcl procedures defined in the scope of core namespaces (namespaces with a call to the \texttt{hbs::Register} procedure).
However, to allow users to define custom utility procedures within cores, procedures with names starting with the floor character (\texttt{\_}) are not treated as core targets.
Listing \ref{lst:edge-detector-hbs} presents an example edge detector core definition.
The core path is \texttt{vhdl::simple::edge-detector}.
The core has three targets: \texttt{src}, \texttt{tb-sync}, \texttt{tb-comb}, and one utility procedure \texttt{\_tb}.
The \texttt{\_tb} procedure was defined to share calls common for testbench targets \texttt{tb-sync} and \texttt{tb-comb}.
Moreover, all target procedures are also regular Tcl procedures.
Such an approach allows for calling them in arbitrary places.
The \texttt{\_tb} procedure calls the \texttt{src} procedure because the \texttt{edge\_detector.vhd} file is required for running the testbench targets.

All targets are represented by a unique target path.
The target path consists of the core path and the target name.
For example, the \texttt{src} target of the edge detector has the following path \linebreak \texttt{vhdl::simple::edge-detector::src}.

\begin{lstlisting}[
  language=Tcl,
  label=lst:edge-detector-hbs,
  caption=Example edge detector core definition.
]
namespace eval vhdl::simple::edge-detector {
  proc src {} {
    hbs::SetLib "simple"
    hbs::AddFile src/edge_detector.vhd
  }
  proc _tb {top} {
    hbs::SetTool "ghdl"
    hbs::SetTop $top
    src
    hbs::SetLib ""
  }
  proc tb-sync {} {
    _tb "tb_edge_detector_sync"
    hbs::AddFile tb/tb_sync.vhd
    hbs::Run
  }
  proc tb-comb {} {
    _tb "tb_edge_detector_comb"
    hbs::AddFile tb/tb_comb.vhd
    hbs::Run
  }
  hbs::Register
}
\end{lstlisting}

\subsection{Testbench targets}

The HBS is capable of automatically detecting testbench targets.
Testbench targets are targets whose names:
\begin{enumerate}
  \item start with \texttt{tb-} or \texttt{tb\_} prefix,
  \item end with \texttt{-tb} or \texttt{\_tb} suffix,
  \item equal \texttt{tb}.
\end{enumerate}

For example, for the edge detector from Listing \ref{lst:edge-detector-hbs}, the \texttt{hbs} program detects testbench targets presented in Listing \ref{lst:edge-detector-tb-targets}.

\begin{lstlisting}[
  label=lst:edge-detector-tb-targets,
  caption=Testbench targets listed for example edge detector core.
]
[user@host tmp]$ hbs ls-tb
vhdl::simple::edge-detector::tb-comb
vhdl::simple::edge-detector::tb-sync
\end{lstlisting}

\subsection{Running targets}
\label{subsec:running-targets}

The HBS allows running any target of registered cores, even if the target itself has nothing to do with the hardware design.
For example, running target \texttt{print} from Listing \ref{lst:print-target-hbs} results with the output shown in Listing \ref{lst:print-target-result}.

\begin{lstlisting}[
  language=Tcl,
  label=lst:print-target-hbs,
  caption=Example print target not related to hardware build.
]
namespace eval core {
  proc print {} {
    puts "Hello!"
  }
  hbs::Register
}
\end{lstlisting}

\begin{lstlisting}[
  label=lst:print-target-result,
  caption=Running print target result.
]
[user@host tmp]$ hbs run core::print
Hello!
\end{lstlisting}

However, in most cases, the user wants to run a target related to the flow of the set EDA tool.
In such a case, instead of manually calling all required tool commands, the user can call the \texttt{hbs::Run} procedure from the core target procedure.

\subsection{Target parameters}
\label{subsec:target-parameters}

As core targets are just Tcl procedures, they can have parameters.
Moreover, parameters can have optional default values.
Additionally, HBS allows providing command-line arguments to the run target.
This is a very convenient feature in build systems.
Listing \ref{lst:target-parameter-hbs} presents a very simplified example.

\begin{lstlisting}[
  language=Tcl,
  label=lst:target-parameter-hbs,
  caption=Example target with an optional parameter.
]
namespace eval core {
  proc target {{stage "bitstream"}} {
    puts "Running until $stage"
    # hbs::Run commented out because this is just an example.
    #hbs::Run $stage
  }
  hbs::Register
}
\end{lstlisting}

The core does not build any hardware design.
However, the example shows how the build stage can be passed from the command line to an EDA tool.
Listing \ref{lst:target-parameter-output} presents output from running the target with different \texttt{stage} parameter values.

\begin{lstlisting}[
  label=lst:target-parameter-output,
  caption=Running print target result.
]
[user@host tmp]$ hbs run core::target
Running until bitstream
[user@host tmp]$ hbs run core::target synthesis
Running until synthesis
\end{lstlisting}

Another practical example of target parameters usage is setting the simulator for testbench target from the command line or changing the top-level module.
What target parameters are used for is limited only by the user's imagination, and Tcl semantics.

\subsection{Target dependencies}

In HBS, targets depend on other targets rather than on cores.
Such an approach allows for fine-grained control of dependencies.

To declare target dependency, the user must call the \texttt{hbs::AddDep} procedure within the target procedure.
The first argument is the dependency path.
The remaining arguments are optional and are forwarded to the dependency target procedure.

To add multiple distinct dependencies, the user must call \texttt{hbs::AddDep} multiple times.
The ability to pass custom arguments to a dependency was evaluated as much more advantageous than the ability to add multiple dependencies with a single \texttt{hbs::AddDep} call.

The \texttt{hbs::AddDep} internally calls the dependency procedure with the provided arguments.
It also tracks dependencies, enabling the generation of a dependency graph.
Within a single flow, each target procedure can be run at most once with a particular set of arguments.
This implies that if multiple target procedures add the same dependency with the same arguments, the dependency procedure is run only once during the first \texttt{hbs::AddDep} call.
To enforce a target procedure rerun, the user can always call the target procedure directly.
However, enforcing target procedure rerun usually is an alert that a regular Tcl procedure shall be used instead of the core target procedure.

\subsection{Code generation}
\label{subsec:code-generation}

Code generation in the hardware design domain is ubiquitous, as it significantly speeds up implementation.
For example, hardware-software co-design system-on-chip projects usually have some tool automatically generating register files \cite{optimized_hw_fw_generation_2020}.
Even in pure FPGA designs, it is common to generate register-transfer-level descriptions from a higher-level programming abstraction \cite{hls_for_fpga, survey_eval_fpga_hls}.
That is why any hardware build system needs to provide as simple mechanism for code generation as possible.

Some existing hardware build systems, for example, FuseSoc, do not allow the direct calling of an arbitrary external program ar arbitrary places during the build process.
Instead, the user has to define so-called generators \cite{fusesoc_generators}.
Only then can the user call the generator within core definitions.
However, such an approach has some drawbacks:
\begin{enumerate}
  \item
  The generator call requires an extra layer of indirection.
  Generators are defined in different places than where they are used, which decreases the readability of the description.
  \item
  The generator call syntax does not resemble shell command call syntax.
  Generators are usually regular applications that can be executed in a shell.
  Calling a generator within the build system using syntax similar to shell seems natural.
\end{enumerate}

In HBS, there is no formal concept of a generator.
Anything can be a generator, as generators are just regular Tcl procedures.
This means that generators can be target procedures (tracked by the dependency system) or core internal Tcl procedures (not tracked by the dependency system).

Listing \ref{lst:hbs-generator-dep} presents an example of calling an external code generator tracked by the dependency system.
In actual use, the call to the shell's echo command would be replaced with a call to the proper code-generation program.
Calls to the \texttt{hbs::AddFile} are commented out because no EDA tool was set.

\begin{lstlisting}[
  language=Tcl,
  label=lst:hbs-generator-dep,
  caption=Example of HBS generator tracked by dependency system.
]
namespace eval core {
  proc top {} {
    hbs::AddDep generator::gen "foo"
    puts "Adding file top.vhd"
    # hbs::AddFile top.vhd
  }
  hbs::Register
}
namespace eval generator {
  proc gen {name} {
    exec echo "Generating $name.vhd" >@ stdout
    puts "Adding file $name.vhd"
    # hbs::AddFile "$name.vhd"
  }
  hbs::Register
}
\end{lstlisting}

Listing \ref{lst:hbs-generator-no-dep} presents how to achieve the same result without tracking the generator as a dependency.
This task is even more straightforward, as the user can call an external generator program directly in the target procedure.
\begin{lstlisting}[
  language=Tcl,
  label=lst:hbs-generator-no-dep,
  caption=Example of HBS generator not tracked by dependency system.
]
namespace eval core {
  proc top {} {
    exec echo "Generating foo.vhd" >@ stdout
    puts "Adding file foo.vhd"
    # hbs::AddFile "foo.vhd"

    puts "Adding file top.vhd"
    # hbs::AddFile top.vhd
  }
  hbs::Register
}
\end{lstlisting}

\subsection{Hardware Build System Commands}
\label{subsec:commands}

The implemented \texttt{hbs} program provides the user with 14 commands.
Listing \ref{lst:hbs-help-msg} presents \texttt{hbs} help message.
Describing particular commands is considered out of scope of the article, as it would not bring any substantive value.
The commands are listed solely to present the maturity and versatility of the implemented build system.
None of the existing indirect-abstract approach hardware build systems support all features present in HBS.
However, some have features missing in HBS, for example, SiliconCompiler, HAMMER, or Hog.

\begin{lstlisting}[
  label=lst:hbs-help-msg,
  caption=HBS help message.
]
Usage

  hbs <command> [arguments]

The command is one of:

  help        Print help message
  doc         Show documentation for cores
  dump        Dump info about cores in Tcl dictionary format
  dump-json   Dump info about cores in JSON format
  graph       Output dependency graph for given target
  info        Show information on hbs Tcl symbol or EDA tool
  ls-cores    List cores found in .hbs files
  ls-targets  List targets for given core
  ls-tb       List testbench targets
  run         Run given target
  dry-run     Run given target without executing and
              evaluating commands
  test        Run testbench targets
  version     Print hbs version
  where       Print where given cores are defined

Type 'hbs help <command>' to obtain more information about
particular command.
\end{lstlisting}

\section{Direct-Tcl and indirect-abstract approaches comparison}
\label{sec:comparison}

This section aims to characterize and compare direct-Tcl and indirect-abstract approaches for hardware build systems.
The goal of the comparison is not to emerge the best hardware build system, but to present differences between build systems utilizing the direct-Tcl approach (build system code executed directly by EDA tools during the design build flow) and the ones utilizing the indirect-abstract approach (build system generates Tcl script which is later run by a proper EDA tool).
The best choice of the hardware build system depends on the project or user requirements.

Comparing hardware build systems is not straightforward.
There are no objective numerical metrics that can be used for the build systems comparison.
One such metric could be build time.
However, in the case of hardware designs, the share of time consumed by EDA tools is orders of magnitude greater than the time overhead introduced by the build system code execution.
This is why the build system execution time overhead is not taken into consideration at all.

Some generic software evaluation metrics could be used, such as numerical or memory complexity.
However, the memory overhead introduced by build systems is negligible compared to the memory utilization of EDA tools.
The numerical complexity might be important for the build system developers, but it is rather irrelevant for build system users.

This section provides a comparative analysis of direct-Tcl and indirect-abstract hardware build systems from a practical perspective.
The qualitative assessment examines key operational attributes, including abstraction level, dependency overhead and its potential implications, internal code structure, constraints handling, handling EDA tools without Tcl interface, arbitrary external program calling, extra features, implementation complexity and learning difficulty.

Table \ref{tab:complexity} presents some numerical and functional features for the analyzed hardware build systems.
The ``$\geq N$'' signature means at least $N$.
For build systems implemented in Python, the number of dependencies was analyzed using the \texttt{pipreqs} \cite{pipreqs_github} tool.
For build systems implemented in Rust, the number of dependencies was extracted from the \texttt{Cargo.toml} files.
For flgen and bazel\_hdl\_rules, the number of dependencies was analyzed manually.
Determining the exact number of dependencies (except for HBS) was not straightforward, as there might be additional runtime dependencies.
Hence, the need for the greater-than-equal sign.

The theory of operation of both approaches was analyzed in section \ref{sec:approaches}, and is not included again in this section.

\begin{table}[]

\centering

\caption{Hardware build systems comparison (Y - yes, N - no, P - partially).}
\label{tab:complexity}

\begin{turn}{90}
\small

\begin{tabular}{c|ccccccccccc}
\textbf{Build System}                                                                                  & \textbf{HBS} & \textbf{FuseSoc} & \textbf{SiliconCompiler} & \textbf{HAMMER} & \textbf{Hog}    & \textbf{hdlmake} & \textbf{bender}  & \textbf{orbit}   & \textbf{Blockwork} & \textbf{flgen}   & \textbf{bazel\_rules\_hdl} \\
\hline
\textbf{\begin{tabular}[c]{@{}c@{}}Number of\\ Files\end{tabular}}                                     & 1            & 67               & 552                      & 496             & 99              & 94               & 73               & 211              & 87                 & 42               & 360                        \\
\hline
\textbf{\begin{tabular}[c]{@{}c@{}}Lines of\\ Code\end{tabular}}                                       & 3723         & 6460             & 58543                    & 136679          & 18982           & 8494             & 12026            & 46690            & 8610               & 3621             & 42743                      \\
\hline
\textbf{\begin{tabular}[c]{@{}c@{}}External\\ Dependencies\end{tabular}}                               & 2            & \textgreater{}12 & \textgreater{}37         & \textgreater{}9 & \textgreater{}3 & \textgreater{}5  & \textgreater{}23 & \textgreater{}20 & \textgreater{}16   & \textgreater{}10 & \textgreater{}7            \\
\hline
\textbf{\begin{tabular}[c]{@{}c@{}}Implementation\\ Language\end{tabular}}                             & Tcl          & Python           & Python                   & Python          & Shell/Tcl       & Python           & Rust             & Rust             & Python             & Ruby             & Starlark                   \\
\hline
\textbf{\begin{tabular}[c]{@{}c@{}}Introduced\\ File Extensions\end{tabular}}                          & 1            & 1                & 0                        & 0               & 4               & 1                & 2                & 1                & 0                  & 1                & 0                          \\
\hline
\textbf{\begin{tabular}[c]{@{}c@{}}Introduced\\ File Syntaxes\end{tabular}}                            & 0            & 0                & 0                        & 0               & 1               & 0                & 0                & 0                & 0                  & 1                & 1                          \\
\hline
\textbf{\begin{tabular}[c]{@{}c@{}}Dependency\\ Graph Generation\end{tabular}}                         & Y            & N                & Y                        & N               & N               & N                & Y                & Y                & N                  & N                & N                          \\
\hline
\textbf{\begin{tabular}[c]{@{}c@{}}Build System\\ Code Executed\\ During\\ Design Build\end{tabular}}  & Y            & N                & N                        & N               & P               & N                & N                & N                & N                  & N                & N                          \\
\hline
\textbf{\begin{tabular}[c]{@{}c@{}}Build System\\ Code Executed\\ Directly by\\ EDA Tool\end{tabular}} & Y            & N                & N                        & N               & P               & N                & N                & N                & N                  & N                & N                          \\
\hline
\textbf{\begin{tabular}[c]{@{}c@{}}Tied to Single\\ Revision System\end{tabular}}                      & N            & N                & N                        & N               & Y               & N                & N                & N                & N                  & N                & N                          \\
\hline
\textbf{\begin{tabular}[c]{@{}c@{}}Requires Specific\\ Directory\\ Structure\end{tabular}}             & N            & N                & N                        & N               & Y               & N                & N                & N                & N                  & N                & N                          \\
\hline
\textbf{\begin{tabular}[c]{@{}c@{}}Capable of\\ Building Designs\end{tabular}}                         & Y            & Y                & Y                        & Y               & Y               & Y                & Y                & N                & Y                  & N                & Y                          \\
\hline
\textbf{\begin{tabular}[c]{@{}c@{}}Requires\\ Containerization\end{tabular}}                           & N            & N                & N                        & N               & N               & N                & N                & N                & Y                  & N                & N                          \\
\hline
\textbf{\begin{tabular}[c]{@{}c@{}}Automatic\\ Testbench\\ Running\end{tabular}}                       & Y            & N                & N                        & N               & N               & N                & N                & N                & N                  & N                & N                          \\
\hline
\textbf{\begin{tabular}[c]{@{}c@{}}Distributed \\ execution of\\ compilation flows\end{tabular}}                & N    & N                & Y                & N               & N               & N               & N                & N                & N                & N                & N
\end{tabular}

\end{turn}


\end{table}

\subsection{Abstraction level}

The direct-Tcl and indirect-abstract approaches present entirely different paradigms for abstraction over EDA tool interfaces.

In the direct-Tcl approach, the abstraction layer over EDA tool interfaces is limited to the common, primary actions.
For example, in HBS, the following actions constitute the common abstraction layer over multiple EDA tools:
\begin{enumerate}
  \item
    Target device setting.
    All synthesis EDA tools require information on the target device.
    This is why setting the device became part of the common abstraction layer.
  \item File addition - this includes support for adding files of all formats supported by a given EDA tool.
  \item Library setting - setting HDL file library.
  \item
    HDL standard setting - setting HDL file standard revision.
    The build system has to manage this because some tools can not analyze different design units with different standard revisions, for example, nvc  \cite{gasson_nvc}.
    In such a case, the build system must decide which common standard revision to use for analyzing all HDL files.
  \item Dependency specification - this is the core feature of any build system.
  \item Generics/parameters setting - configuring parametric designs must be an inherent feature of any hardware build system.
  \item Design top module setting - all EDA tools performing simulation or synthesis require information about the top module.
  \item
    Exit severity setting - testbenches are designed to exit with an error code when a specific severity message occurs.
    This is independent of the simulator being used, hence it should be hidden under the build system abstraction.
\end{enumerate}

In the direct-Tcl approach, the build system intentionally exposes EDA tool-specific quirks so that the user has to resolve them explicitly.
This is usually achieved with tool guard clauses.
Listing \ref{lst:if-guard} presents an example of such a guard.
The snippet presents a helper testbench procedure definition for the AMBA 5 APB crossbar.
If the tool is set to the nvc simulator, an extra argument "--dump-arrays" must be passed to the simulator to dump array signals to the waveform file.
This can be seen as a peculiarity of the nvc simulator.
\begin{lstlisting}[
  label=lst:if-guard,
  caption=An example of tool guard clause in HBS (direct-Tcl approach).
]
proc _tb {top} {
  hbs::AddDep vhdl::amba5::apb::checker::src
  hbs::AddDep vhdl::amba5::apb::bfm::src
  hbs::AddDep vhdl::amba5::apb::mock-completer::src
  if {$hbs::Tool eq "nvc"} {
    hbs::AddPreSimCb hbs::SetArgSuffix "--dump-arrays"
  }
  hbs::SetTop $top
  hbs::AddDep $hbs::ThisCorePath\::src
}
\end{lstlisting}

In the indirect-abstract approach, the main purpose of the abstraction layer is to provide a great, unified abstraction over EDA tools.
As such, the abstraction layer covers many more aspects than the abstraction layer in the direct-Tcl approach.
However, maximizing the abstraction layer is not the goal of the indirect-abstract approach; the goal is to unify the user experience.

Instead of exposing EDA tools peculiarities to the user, the indirect-abstract approach tries to hide them under the abstraction layer.
For example, the nvc peculiarity of not dumping arrays signals by default, in the indirect-abstract approach is solved in such a way, that build systems automatically add the "--dump-arrays" argument when simulation is run.

\subsection{Dependency overhead and its potential implications}

The term ''dependency'' in the following context means a program, library, or operating system package that must be explicitly installed by the user to utilize all features of the analyzed build system.
Any tools installed automatically alongside EDA tools are not treated as dependencies.

Hardware build systems representing the direct-Tcl approach tend to have fewer dependencies.
For example, the implemented HBS has only the following two dependencies:
\begin{enumerate}
  \item
  Tcl shell (mandatory) - required to run the core logic of the build system.
  \item
  Graphviz (optional) - required for generating a core dependency graph in a graphical format.
\end{enumerate}
None of the existing hardware build systems representing the indirect-abstract approach, capable of similar features, has so few dependencies.

The explanation for the lower number of dependencies in build systems that use the direct-Tcl approach is straightforward.
They rely on libraries and programs that are installed during EDA tool installation.
On the other hand, build systems utilizing the indirect-abstract approach often rely on dependencies installed manually via package managers.

Nevertheless, a pure number of dependencies might not be a good metric for comparison.
A project could have a single dependency that suffers from multiple issues, and another project could have hundreds of well-maintained dependencies.

What might be more important from the user's point of view is stability.
A build system is expected to work across many systems with a wide range of Long Term Support.
The ease of initial setup of the build system might also be relevant to the user.

The direct-Tcl approach with fewer dependencies is probably less susceptible to stability issues.
However, overall stability depends on multiple factors, such as server stability, OS stability, network reliability, CI/CD complexity, etc.

\subsection{Internal code structure}
\label{subsec:code-struct}

The internal code structure was analyzed in terms of the number of files, lines of code, introduced file syntaxes and extensions, and supported features.
The following types of files were excluded from analysis: author files, news files, contributors files, license files, readme files, tests files, and example files.
The goal of the analysis is to provide raw facts.

In Table \ref{tab:complexity}, the introduced file extensions are the number of custom file extensions used by a build system.
The introduced file syntaxes are the number of custom syntaxes used by a build system.
For example, FuseSoc scans for files with the \texttt{.core} extension.
However, the \texttt{.core} files have a standard YAML syntax.
This is why the number of introduced file extensions equals 1, but the number of introduced file syntaxes equals 0.

An extra context has to be introduced to the table, as different tools support different features and EDA tools.
Direct-Tcl build systems (except HBS) are not included in the table.
This is because they support only one EDA tool, and they do not support automatic testbench detection and dependency graph generation.
OSVVM-Script is not included because it targets only simulation tools.

The FuseSoc build tool internally uses the external Edalize \cite{edalize_github} library to interact with EDA tools.
Including Edalize, the number of files for FuseSoc equals 215, and the number of lines of code equals 19788.

The total number of lines of code of any hardware build system depends heavily on the number of supported EDA tools.
FuseSoc supports the greatest number of tools, almost 40.
HBS currently supports the following seven tools: AMD Vivado, AMD xsim, GHDL \cite{gingold_ghdl}, GOWIN IDE, Intel/Altera Quartus, nvc, Questa.
The most complex one is Vivado, which full support requires roughly 190 lines of code.
However, supporting the nvc simulator required roughly 170 lines of code.
Assuming that HBS will one day support 40 EDA tools, the number of lines of code should remain below 10000.

Tools following the direct-Tcl approach tend to have more monolithic code structure, fewer files, and fewer lines of code.
The code for handling EDA tools is often placed in a single file.
The lower number of lines of code results from the direct access to the EDA tool commands.
There is no need for wrapper code.
They prefer shared global variables which makes build logic easier to follow but also more bug-prone.

Tools following the indirect-abstract approach tend to have a more modular code structure.
They have more files and a greater number of lines of code.
The code for handling specific EDA tools is organized into modules in separate files.
The encapsulation makes the code more bug-resistant, but the build logic is slightly harder to follow.

\subsection{Constraints handling}
\label{subsec:constr-handling}

Building any hardware design (FPGA or ASIC) without constraints is impossible.
Constrains usually define clocks, path delays, timing relations, and physical properties, for example, pin locations, drive strength or voltage level.
Any hardware build system must support various constraint types and scenarios.
The goal of this subsection is to present differences in constraint handling from the user perspective of the build system.

In the case of a simple constraints scenario with global constraints, there is no difference between the direct-Tcl and indirect-abstract approaches.
The user simply defines constraints in a supported file format (e.g., .sdc, .xdc, or .tcl) and adds the file to the project or build flow.
Based on the file extension, build systems decide how to handle the file.
However, there is a noticeable difference when a project requires constraints scoped to modules and cells, or when constraints are added optionally.

\section*{Scoping constraints to module}
\label{lab:scoping-constr-to-module}

Let us analyze a simple example.
The engineer implements a module and wants to scope some constraints to this module.
As an example, a two-stage flip-flop clock domain crossing synchronizer is used.
The source code of the synchronizer is not relevant to the analysis.
The false path is set as a constraint, although the max delay is also frequently used.

Listing \ref{lst:fusesoc-synchronizer} presents the synchronizer core definition in FuseSoc, which represents the declarative indirect-abstract subcategory.

\begin{lstlisting}[
  label=lst:fusesoc-synchronizer,
  caption=Cross domain crossing synchronizer core definition in FuseSoc.
]
CAPI=2:

name: vendor:library:synchronizer

filesets:
  src:
    files:
      - synchronizer.vhd
    file_type: vhdlSource-2008
  vivado_constr:
    files:
      - vivado-constr.xdc: {file_type: xdc}
  tcl:
    files:
      - constr.tcl: {file_type: tclSource}

targets:
  default:
    filesets:
      - src
      - "tool_vivado? (vivado_constr)"
      - tcl
\end{lstlisting}

Listing \ref{lst:fusesoc-vivado-constr} presents the content of the \texttt{vivado-constr.xdc} file, and Listing \ref{lst:fusesoc-constr-tcl} presents the content of the \texttt{constr.tcl} file.

\begin{lstlisting}[
  label=lst:fusesoc-vivado-constr,
  caption=Content of the \texttt{vivado-constr.xdc} file.
]
set_false_path -to [get_pins s_0_reg[*]/D]
\end{lstlisting}

\begin{lstlisting}[
  language=Tcl,
  label=lst:fusesoc-constr-tcl,
  caption=Content of the \texttt{constr.tcl} file for FuseSoc.
]
if {[string match "Vivado*" [version]]} {
  set_property SCOPED_TO_REF Synchronizer \
    [get_files vivado-constr.xdc]
} else {
  error \
    "Synchronizer entity misses constraint file for your EDA tool"
}
\end{lstlisting}

In FuseSoc, to accomplish the task, the user must provide four files:
\begin{enumerate}
  \item \texttt{synchronizer.vhd} - synchronizer HDL description,
  \item \texttt{vivado-constr.xdc} - false path Vivado constraint for synchronizer,
  \item \texttt{constr.tcl} - Tcl script scoping constraint to synchronizer module.
  \item \texttt{synchronizer.core} - synchronizer core definition.
\end{enumerate}
The number of required files (4) is greater than the number of logical elements (3).
The user describes the module in HDL, defines the core for the build system, and adds extra constraints.
The number of logical elements equals three, and the number of files equals four.
The logic of the core definition is more complex to follow, as the user has to switch between files for a very short snippet of code.
This may lead to a decrease in productivity \cite{Meyer2014}.

Listing \ref{lst:hbs-synchronizer} presents the same core definition but in the direct-Tcl approach.
The number of required files is three and equals the number of logical elements.
The \texttt{constr.tcl} file could be eliminated because the code of the build system is executed directly by the EDA tool.
Even the error message is improved.
It is possible to report which EDA tool the constraint file definition is missing for.
Again, this is possible because the build system code is executed directly by the EDA tool.

In the case of the indirect-abstract approach, build scripts are prepared before the EDA tool is run.
Within the \texttt{constr.tcl} script, there is no access to the build system context or variables, as the build system code execution is already finished.

\begin{lstlisting}[
  language=Tcl,
  label=lst:hbs-synchronizer,
  caption=Cross domain crossing synchronizer core definition in HBS (direct-Tcl approach).
]
namespace eval vendor::library::synchronizer {
  proc src {} {
    hbs::AddFile synchronizer.vhd

    if {$hbs::Tool == "vivado-prj"} {
      hbs::AddFile vivado-constr.xdc
      set_property SCOPED_TO_REF Synchronizer \
        [get_files vivado-constr.xdc]
    } else {
      error \
        "Synchronizer entity misses constraint file for $hbs::Tool"
    }
  }

  hbs::Register
}
\end{lstlisting}

In the case of the programmable indirect-abstract subcategory, the solution depends on the tool.
For example, in hdlmake, the intermediate Tcl script is also required to scope a constraint file to a module.
The same script is also required for the SiliconCompier.
Hog, for example, instead of a Tcl script, requires a custom constraint list file (a file with the \texttt{.con} extension) with paths to constraint files and their property declarations.
Moreover, the only way to opt out of the default constraints is to comment out the constraint lines in the constraint list file.
To the author's knowledge, none of the existing indirect-abstract build systems utilizes the concept of direct injection of Tcl code into the build script.
They all need one additional file (a Tcl script or a tool-specific file) for scoping constraints.

\section*{Scoping constraints to cells and optional constraints}

Let us analyze a more complex example.
Namely, the Advanced Peripheral Bus Clock Domain Crossing bridge (APB CDC bridge).
Such a bridge requires some timing constraints due to the clock domain crossing.
There are a few possible ways to define valid constraints for the APB CDC bridge.
The core designer may want to provide some default constraints.
However, the user should also be able to provide their own constraints.

One possible constraint strategy is to set a false path for handshake signals and a maximum delay for bus signals crossing domains.
The max delay value should equal the period of the destination clock.
The false path constraints can be safely scoped to the APB CDC bridge module.
However, because there may be multiple APB CDC bridges in the design crossing between various clock domains, the max delay constraints should be scoped to cells.
An alternative solution is to pick the shortest clock period as the value for the max delay constraint for all bridges.
However, such an approach leads to design overconstraint and imposes unnecessary requirements on the router.

Listing \ref{lst:apb-cdc-bridge} shows how APB CDC bridge constraints are handled in the HBS, which represents the direct-Tcl approach.
The code is concise and located within a single procedure.
The user can pass the \texttt{-no-constr} flag to the target to omit adding the default constraints.
The flag is checked with a simple if statement.
Moreover, if an invalid argument is passed, an error with a precise message is returned.
The \texttt{apb-cdc-bridge-ref.xdc} and \texttt{apb-cdc-bridge-cell.tcl} constraint files are not presented as their content is irrelevant.
The first one has constraints scoped to the module; the second one has code that applies constraints scoped to cells.

\begin{lstlisting}[
  language=Tcl,
  label=lst:apb-cdc-bridge,
  caption=Advanced Peripheral Bus Clock Domain Crossing bridge target definition in HBS (direct-Tcl).
]
# Use -no-constr flag to omit adding the default constraints.
proc src {args} {
  hbs::AddDep vhdl::amba5::apb::pkg::src
  hbs::SetLib "lapb"
  hbs::AddFile cdc-bridge.vhd

  set argCount [llength $args]
  if {$argCount == 1} {
    set arg [lindex $args 0]
    if {$arg eq "-no-constr"} {
      return
    } else {
      error "$hbs::ThisTargetPath: invalid arg '$arg', " \
            "valid args are: '-no-constr'"
    }
  } elseif {$argCount > 1} {
    error \
      "$hbs::ThisTargetPath: invalid arg count $argCount, " \
      "expected at most 1"
  }

  if {$hbs::Tool eq "vivado-prj"} {
    hbs::AddFile vivado/apb-cdc-bridge-ref.xdc
    set_property SCOPED_TO_REF APB_CDC_Bridge \
      [get_files apb-cdc-bridge-ref.xdc]
    hbs::AddFile vivado/apb-cdc-bridge-cell.tcl
    set_property USED_IN_SYNTHESIS false \
      [get_files apb-cdc-bridge-cell.tcl]
  } else {
    error "$hbs::ThisCore misses constraints for $hbs::Tool"
  }
}
\end{lstlisting}

How applying the same constraints in the indirect-abstract approach looks depends on the tool subcategory.
For example, in FuseSoc, which represents the declarative subcategory, applying the module and cell-scoped constraints would be similar to what was presented in listings \ref{lst:fusesoc-synchronizer} and \ref{lst:fusesoc-constr-tcl} for the CDC synchronizer.
Again, one more Tcl file would be required.
The optionality of the constraints would be handled using the flags mechanism \cite{fusesoc_flags}.
While the flag mechanism allows conditional inclusion of files, it does not allow arbitrary actions to be executed due to its declarative nature.

In hdlmake, the same core description would require two \texttt{Manifest.py} files.
The first one with constraints included, and the second one without.
The user would pick a proper \texttt{Manifest.py} file when declaring dependency on the core.
With just a single boolean parameter, this approach does not seem complex.
However, with multiple parameters, the number of \texttt{Manifest.py} files would equal the number of possible combinations.

In the programmable subcategory of the indirect-abstract approach, the complexity of constraint handling is lower.
In this subcategory, the solution is very similar to the one present in the direct-Tcl approach.
However, instead of directly injecting the Tcl code into the generated Tcl build script, the user must place it in a separate Tcl file.
There is only minor overhead resulting from:
\begin{enumerate}
  \item build system programming language Application Programming Interface (API) call,
  \item build logic fragmentation due to the requirement of putting the Tcl code in a separate file.
\end{enumerate}

\subsection{Handling EDA tools without Tcl interface}

Not all EDA tools have a Tcl interface.
For example, free and open-source simulators, for example, Verilator \cite{snyder_verilator} or GHDL, have a shell interface.
The user simply calls programs within the shell providing appropriate arguments.

Handling EDA tools without the Tcl interface in the direct-Tcl approach is straightforward.
The Tcl language makes it easy to call any external program using the \texttt{exec} command.
Listing \ref{lst:ghdl-elab} presents how the elaboration stage for the GHDL simulator is implemented in HBS.
The command to be executed is constructed directly as a string to be passed to the shell.
The code related to the working directory change and error checking takes more lines than the code creating the actual command.
The \texttt{hbs::Exec} procedure is a simple wrapper around the Tcl \texttt{exec} command, which allows printing the command to the standard output in the dry-run mode.

\begin{lstlisting}[
  language=Tcl,
  label=lst:ghdl-elab,
  caption=GHDL elaboration stage implementation in HBS (direct-Tcl approach).
]
proc elaborate {} {
  set workDir [pwd]
  cd $hbs::RunTargetBuildDir

  set cmd "ghdl -e $hbs::ArgPrefix --std=[hbs::ghdl::std] " \
          "--workdir=work $hbs::ghdl::libs $hbs::ArgSuffix $hbs::Top"
  set err [hbs::Exec $cmd]
  if {$err} {
    hbs::panic "$hbs::Top elaboration failed with exit status $err"
  }

  cd $workDir
}
\end{lstlisting}

In the indirect-abstract approach, handling EDA tools without the Tcl interface is implemented differently.
In the indirect-abstract approach, the build system code is run before the build code.
This implies that the build system code must generate the build code that is only later run.
The build code is usually generated in the form of a Makefile or shell script.
Compared to the direct-Tcl approach, there are two essential differences:
\begin{enumerate}
  \item The generated build code script can be easily reused without rerunning the build system code.
  \item
    The build system code logic might be slightly harder to follow.
    This is because the data for the script templates is usually not prepared in a linear way that corresponds to the build flow.
    Whereas in the direct-Tcl approach, the logic must be linear because the build code is executed during the build system code execution.
\end{enumerate}

\subsection{Arbitrary external program calling}
\label{subsec:external-program-calling}

Arbitrary external program calling is a very useful feature of modern hardware build systems.
The word "arbitrary" means two things:
\begin{enumerate}
  \item arbitrary program - the user should be able to call any program available on the operating system,
  \item arbitrary place - the user should be able to call programs in any place within the design tool flow.
\end{enumerate}

A practical example of an arbitrary program is a code generator.
Code generation in the hardware design domain is ubiquitous, as it significantly speeds up implementation.

A practical example of an arbitrary place is report analysis.
A user may want to analyze the synthesis stage results before starting the implementation stage.
However, the user may want to implement the reports parsing logic in a Perl or Python script.
In such a case, there must be a way to call the custom script within the Tcl build script after the synthesis.

Meeting the two arbitrariness requirements (program and place) in the direct-Tcl approach is straightforward.
This is because of the following reasons:
\begin{enumerate}
  \item Tcl has built-in \texttt{exec} command for executing external programs,
  \item the build system code is executed directly by the EDA tool,
  \item the build system code is executed during the design build flow.
\end{enumerate}

Listing \ref{lst:hbs-exec-after-synth} presents an example of how to execute a command after the synthesis step in HBS.

\begin{lstlisting}[
  label=lst:hbs-exec-after-synth,
  caption=Post synthesis callback example in HBS (direct-Tcl).
]
proc myPostSynthCb {} {
  set err [catch {eval exec "./analyze-synth-reports.py"} output]
  if {$output ne ""} {
    puts "synthesis results are not satisfactory: $output"
    exit 1
  }
}

hbs::AddPostSynthCb myPostSynthCb
\end{lstlisting}

\begin{figure}[!t]
\centering
\includegraphics[scale=0.35]{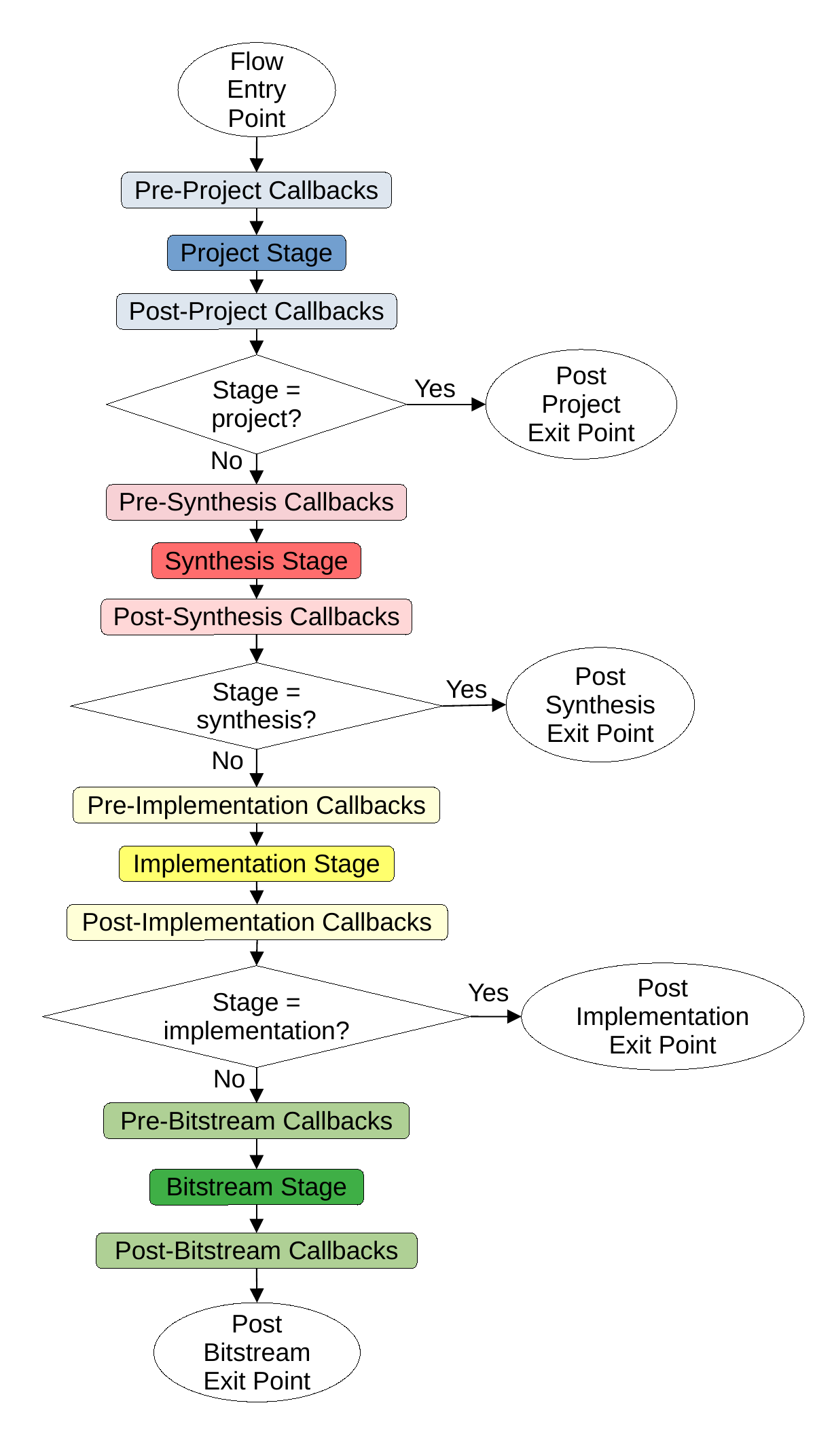}
\caption{Vivado project mode tool flow.}
\label{fig:vivado-flow}
\end{figure}

In HBS, each stage has optional pre- and post-callbacks.
Figure \ref{fig:vivado-flow} shows the structure of the tool flow for the Vivado project mode.

There are four stages: project, synthesis, implementation, and bitstream.
The number of callbacks is unlimited.
Callbacks in any pre- or post-stage are executed in the order users add them.
The user adds a callback by calling a dedicated HBS API procedure.
For example, to add a post-synthesis callback, the user can call the \texttt{hbs::AddPostSynthCb} procedure.
Callbacks can be also added with custom argument values.

While most existing indirect-abstract hardware build systems allow calling any third-party programs, the tools representing the declarative subcategory, do not allow calling them in arbitrary places.
For example, in FuseSoc, the user has to define so-called generators \cite{fusesoc_generators}.
Only then can the user call the generator within core definitions.
However, such an approach has some drawbacks:
\begin{enumerate}
  \item
  The generator call requires an extra layer of indirection.
   Generators are defined in different places from where they are used, which reduces the readability of the description.
  \item
  The generator call syntax does not resemble shell command call syntax.
  Generators are usually regular applications that can be executed in a shell.
  Calling a generator within the build system using syntax similar to shell seems natural.
  \item
  Generators can be executed only before the EDA tool flow.
\end{enumerate}

In the case of build systems representing the programmable indirect-abstract subcategory, meeting the two arbitrariness requirements is also straightforward.
However:
\begin{enumerate}
  \item it requires some Python wrapper code,
  \item the code must be placed in a separate script file.
\end{enumerate}

Listing \ref{lst:siliconcompiler-exec-after-synth} shows how to add a post-synthesis script in SiliconCompiler.
There are dedicated Python API functions, so the procedure is very concise.
Compared to the direct-Tcl approach, there is one more indirection layer, as the actual Tcl code must be placed in a separate file, which leads to the build logic fragmentation.

\begin{lstlisting}[
  label=lst:siliconcompiler-exec-after-synth,
  caption=Post synthesis callback example in SiliconCompiler (indirect-abstract).
]
SynthesisTask.find_task(project).add_postscript(
  "path/to/post-synth-script.tcl"
)
\end{lstlisting}

Listing \ref{lst:hdlmake-exec-after-synth} presents how to execute a command after the synthesis stage in hdlmake.
The snippet is a part of a \texttt{Manifest.py} file.
The hdlmake, like SiliconCompiler, requires the code to be placed in a separate Tcl file.
Moreover, hdlmake closes the project after every step.
If the evaluated code requires access to the build context, for example, synthesis results, the user must explicitly open the project in the script.

\begin{lstlisting}[
  label=lst:hdlmake-exec-after-synth,
  caption=Post synthesis program execution example in hdlmake (indirect-abstract).
]
syn_post_synthesize_cmd = "$(TCL_INTERPRETER) user_tcl_script.tcl;"
\end{lstlisting}

\subsection{Extra features support}
\label{subsec:extra-features-support}

Modern build systems are expected to do much more than produce the final production or programming files.
In terms of hardware build systems, for example, the following extra features are desired:
\begin{enumerate}
  \item automatic testbench targets detection,
  \item parallel testbench running,
  \item core dependency graph generation,
  \item distributed development and distributed execution of compilation flows.
\end{enumerate}

None of the existing direct-Tcl hardware build systems supports feature number 4.
Moreover, HBS is the only direct-Tcl build system supporting features 1, 2, and 3.

In the indirect-abstract class, only SiliconCompiler, bender, and orbit support feature number 3.
The SiliconCompiler is the only existing hardware build system with built-in support for distributed development and distributed execution of compilation flows.

SiliconCompiler and HAMMER have the most extensive support for ASIC and VLSI designs.
Both were designed from the ground up for that purpose and used for commercial projects in this domain.
However, both of them have the greatest internal complexity.

It is also important to note that:
\begin{itemize}
  \item
    The bender is only a dependency management tool.
    It provides a way to define dependencies among cores, execute unit tests, and verify that the source files are valid input for various simulation and synthesis tools.
    However, it does not run the EDA tool for building designs.
  \item
    The flgen is just a file list generator.
    It does not run EDA tools for building designs.
\end{itemize}

\subsection{Extra features implementation complexity}

While hardware build system users are mostly interested in extra-feature support (subsection \ref{subsec:extra-features-support}), hardware build system developers might also be interested in the implementation complexity of extra features.

Implementing extra, non-build-related features in the direct-Tcl approach is more complex than in the indirect-abstract approach.
This state results directly from the use of the Tcl language.
While multi-threading in Tcl is possible, the Tcl was not designed with multi-threading as one of its primary goals.
This explicitly impacts support for parallel testbench execution or distributed compilation.
Moreover, Tcl has fewer third-party libraries that could help implement additional features.

Yet another important factor is the fact that direct-Tcl build systems are executed by the Tcl interpreters provided by the EDA tools.
Each version of the EDA tool is shipped with a fixed version of the Tcl interpreter.
Modern third-party Tcl libraries might require a higher version of the Tcl interpreter, which might potentially block them from being used within EDA tools.
In the indirect-abstract approach, the build system infrastructure (tools and libraries) is managed independently from the EDA tools.

The direct-Tcl approach tools tend to have more code related to the extra features implementation because any feature requiring parallelism has to be implemented asynchronously using the \texttt{filevent} and \texttt{vwait} mechanisms.
Moreover, the code has a non-linear structure and is spread across multiple locations.
An alternative approach to mitigate the mentioned deficiencies in the direct-Tcl approach is to implement the extra features in a wrapper program or script written in another programming language.
However, this leads to splitting the build system's code logic across multiple languages, which might increase cognitive complexity.
Hardware build systems that follow the indirect-abstract approach are usually free of the mentioned drawbacks, as they only generate Tcl code and are entirely written in languages more suitable for implementing such extra features.

\subsection{Learning difficulty}

The direct-Tcl approach might be harder to learn for less experienced engineers.
There are two reasons for such a state being:
\begin{enumerate}
  \item The Tcl language is not considered to be the most user-friendly.
  \item The user must be familiar with the EDA tool commands in the case of less common tasks, for example, working with block designs or generating vendor cores using the Tcl interface instead of the IP wizard.
\end{enumerate}

The indirect-abstract approach has a lower entry point.
In the case of the declarative subcategory, the user must only know basic formats, such as YAML or TOML, and a few build system keywords.
In the case of the programmable subcategory, the language of choice is Python, which is commonly considered easier to learn than Tcl.

It is noteworthy that, in contemporary contexts, the complexity associated with programming languages and EDA tools is effectively mitigated through the application of artificial intelligence (AI) techniques \cite{Kazemitabaar2024Novice, Pearce2023Challenges}.

\section{Summary}
\label{sec:summary}

Table \ref{tab:summary} summarizes characteristics of direct-Tcl and indirect-abstract approaches for hardware build systems.
\begin{table}[]

\centering
\small

\caption{Direct-Tcl and indirect abstract hardware build system approaches comparison summary.}
\label{tab:summary}

\hspace*{-3cm}
\begin{tabular}{c|l|l}

\textbf{Approach}                                                                               & \multicolumn{1}{c|}{\textbf{direct-Tcl}}                                                                                                                                                                                                                                             & \multicolumn{1}{c}{\textbf{indirect-abstract}}                                                                                                                                                                                                                                                                                                 \\ \hline
\textbf{\begin{tabular}[c]{@{}c@{}}Theory of\\ operation\end{tabular}}                          & \begin{tabular}[c]{@{}l@{}}The build system code is executed during\\ the design build flow. The build system code\\ is executed directly by the EDA tool. The\\ build system code and build code reside\\ in the same domain.\end{tabular}                                          & \begin{tabular}[c]{@{}l@{}}The build system code is executed before\\ the design build flow. A build system generates\\ Tcl or shell script that is only later run by the\\ EDA tool. The build system code and build\\ code reside in different domains.\end{tabular}                                                                         \\ \hline
\textbf{\begin{tabular}[c]{@{}c@{}}Dependency\\ overhead\\ and its\\ implications\end{tabular}} & \begin{tabular}[c]{@{}l@{}}Fewer dependencies. Initial environment\\ setup requires fewer installations. Lower\\ probability of stability problems on Long\\ Term Support distributions.\end{tabular}                                                                                & \begin{tabular}[c]{@{}l@{}}More dependencies. Initial environment\\ setup requires more actions. Slightly higher\\ probability of stability problems on Long\\ Term Support distributions.\end{tabular}                                                                                                                                        \\ \hline
\textbf{\begin{tabular}[c]{@{}c@{}}Internal\\ code structure\end{tabular}}                      & \begin{tabular}[c]{@{}l@{}}Monolithic. The code containing logic\\ for EDA tools handling is placed within\\ a single file. Prefers shared global\\ variables which makes build logic easier\\ to follow but also bug-prone.\end{tabular}                                            & \begin{tabular}[c]{@{}l@{}}Modular. The code for handling particular\\ EDA tools is placed within modules in \\ separate files. The encapsulation makes the\\ code more bug-resistant, but the build logic\\ is slightly harder to follow.\end{tabular}                                                                                        \\ \hline
\textbf{\begin{tabular}[c]{@{}c@{}}Abstraction\\ level\end{tabular}}                            & \begin{tabular}[c]{@{}l@{}}Low. The common abstraction layer is\\ limited to primary tasks. For example, file\\ faddition, library setting, generic/parameter\\ setting. Intentionally exposes EDA tool\\ peculiarities to the user.\end{tabular}                                    & \begin{tabular}[c]{@{}l@{}}High. The common abstraction layer tries to\\ cover all areas and hide the peculiarities of\\ different EDA tools. Makes porting a design\\ to a different EDA tool easier.\end{tabular}                                                                                                                            \\ \hline
\textbf{\begin{tabular}[c]{@{}c@{}}Constraints\\ handling\end{tabular}}                         & \begin{tabular}[c]{@{}l@{}}Simple. As the build code is executed\\ directly by the EDA tool during the build\\ system code execution, there is direct\\ access to all constraint commands.\\ Optional constraints addition can be\\ handled with a simple if statement.\end{tabular} & \begin{tabular}[c]{@{}l@{}}Depends on the subcategory. In the case of\\ the declarative subcategory, some less common\\ constraint scenarios are more complex to\\ achieve and require indirection layers. In the\\ case of the programmable subcategory, \\ constraints handling requires only a small\\ amount of wrapper code.\end{tabular} \\ \hline
\textbf{\begin{tabular}[c]{@{}c@{}}Handling EDA\\ tools without\\ Tcl interface\end{tabular}}   & \begin{tabular}[c]{@{}l@{}}Straightforward. These kinds of tools are\\ directly called using the exec command\\ or catch-eval-exec sequence if error\\ handling is required. Build rerun requires\\ build system code rerun.\end{tabular}                                            & \begin{tabular}[c]{@{}l@{}}Slightly more complex. EDA tools without\\ the Tcl interface are handled by generating\\ Makefile or shell scripts. The generated build\\ code scripts can be easily reused without\\ rerunning the build system code.\end{tabular}                                                                                 \\ \hline
\textbf{\begin{tabular}[c]{@{}c@{}}Arbitrary\\ external\\ program\\ calling\end{tabular}}       & \begin{tabular}[c]{@{}l@{}}Straightforward. User simply calls Tcl\\ exec command if error catching is not\\ required or catch-eval-exec sequence if\\ error catching is required. Programs can\\ be easily called between any EDA tool\\ flow stages.\end{tabular}                   & \begin{tabular}[c]{@{}l@{}}Depends on the core description file format.\\ In the case of markup or data serialization\\ languages, calling external programs is\\ complex. In the case of, for example, Python,\\ calling external programs is simple, although\\ it still requires slightly more code than in Tcl.\end{tabular}               \\ \hline
\textbf{\begin{tabular}[c]{@{}c@{}}Extra  features\\ support\end{tabular}}                      & \begin{tabular}[c]{@{}l@{}}Only HBS supports automatic testbench\\ detection, parallel testbench running, and\\ dependency graph generation. No tool\\ natively supports distributed execution of\\ compilation flows.\end{tabular}                                                  & \begin{tabular}[c]{@{}l@{}}Highly depends on the tool. For example, \\ SiliconCompiler is the only tool natively\\ supporting the distributed execution of\\ compilation flows. Only SiliconCompiler,\\ bender and orbit support core dependency\\ graph generation.\end{tabular}                                                              \\ \hline
\textbf{\begin{tabular}[c]{@{}c@{}}Extra features\\ implementation\\ complexity\end{tabular}}   & \begin{tabular}[c]{@{}l@{}}Higher. Extra features, for example, \\ parallel testbench running, require more \\ code. Moreover, the code is asynchronous\\ and has a non-linear structure. The build\\ system code gets fragmented.\end{tabular}                                      & \begin{tabular}[c]{@{}l@{}}Lower. Build systems representing this\\ approach don't have to be implemented in\\ Tcl. Most of them use Python, which\\ makes implementing extra features easier\\ and does not split the build system code logic.\end{tabular}                                                                                   \\ \hline
\textbf{\begin{tabular}[c]{@{}c@{}}Learning\\ difficulty\end{tabular}}                          & \begin{tabular}[c]{@{}l@{}}Higher. The user must directly write Tcl.\\ Moreover, the user has to be familiar with\\ EDA tools commands in the case of less\\ common tasks, for example, working with\\ block designs.\end{tabular}                                                   & \begin{tabular}[c]{@{}l@{}}Lower. Markup, data serialization, and Python\\ syntaxes are much simpler and easier to read\\ even for newcomers.\end{tabular}

\end{tabular}
\end{table}

The taxonomical data presented in the table can serve as a strategic framework for selecting an optimal hardware build system based on specific project requirements and organizational constraints.
For example, in the case of massive-scale ASIC projects that scale to even thousands of developers and millions of servers, the indirect-abstract approach is currently the only viable solution.
Valid choices might be SiliconCompiler or HAMMER, as both provide the necessary abstraction layers to manage this complexity.
 For small-to-medium projects with a low entry threshold, a pure declarative, indirect-abstract hardware build system might be a good choice, such as FuseSoc.
In the case of tight coupling with Git and CI/CD, Hog might be the best choice.
For medium-to-large projects targeting a single-vendor ecosystem, a direct-Tcl approach (e.g., HBS) is often superior.
The direct-Tcl approach is particularly effective for teams with deep expertise in EDA tool command interfaces, where the steeper learning curve is offset by granular control over the implementation flow.

Beyond selection, the comparative analysis in Table \ref{tab:summary} can provide a roadmap for developing next-generation hardware build systems.
By identifying the inherent bottlenecks and architectural weaknesses of existing approaches, developers can proactively implement mitigation strategies.
Understanding these trade-offs a priori allows for the design of more resilient, scalable hardware build systems that address the limitations of current direct-Tcl and indirect-abstract tools.

\vspace{6pt}






\authorcontributions{Not applicable.}

\funding{This research received no external funding. The APC was funded by the Warsaw University of Technology.}

\institutionalreview{Not applicable.}

\informedconsent{Not applicable.}

\dataavailability{Not applicable.}

\acknowledgments{Not applicable.}

\conflictsofinterest{The author declares no conflict of interest.}


\abbreviations{Abbreviations}{
The following abbreviations are used in this manuscript:

\noindent
\begin{tabular}{@{}ll}
AI & Artificial Intelligence\\
AMBA & Advanced Microcontroller Bus Architecture\\
APB & Advanced Peripheral Bus \\
AMD & Advanced Micro Devices \\
ASIC & Application-Specific Integrated Circuit\\
CDC & Clock Domain Crossing\\
CI/DC & Continuous Integration /  Continuous Deployment\\
EDA & Electronic Design Automation\\
FPGA & Field-Programmable Gate Array\\
GNU & GNU’s not Unix\\
HBS & Hardware Build System\\
HDL & Hardware Description Language\\
IDE & Integrated Development Environment\\
IP & Intellectual Property\\
OS & Operating System\\
VLSI & Very-Large-Scale Integration\\
VLNV & Vendor Library Name Version\\
\end{tabular}
}

\appendixtitles{no} 

\begin{adjustwidth}{-\extralength}{0cm}

\reftitle{References}


\bibliography{main.bib}


%


\PublishersNote{}
\end{adjustwidth}
\end{document}